# *Ab initio* study of edge effect on relative motion of walls in carbon nanotubes

Andrey M. Popov[1,a)], Irina V. Lebedeva[2,b)], Andrey A. Knizhnik[2,3,c)], Yurii E. Lozovik[1,4,d)] and Boris V. Potapkin[2,3]

[1]*Institute of Spectroscopy, Russian Academy of Science, Fizicheskaya Street 5, Troitsk, Moscow Region, 142190, Russia,*

[2]*Kintech Lab Ltd., Kurchatov Square 1, Moscow, 123182, Russia,*

[3]*National Research Centre "Kurchatov Institute", Kurchatov Square 1, Moscow, 123182, Russia,*

[4]*Moscow Institute of Physics and Technology, Institutskii pereulok 9, Dolgoprudny, Moscow Region, 141700, Russia*

Interwall interaction energies of double-walled nanotubes (DWNTs) with long inner and short outer walls are calculated as functions of coordinates describing relative rotation and displacement of the walls using van der Waals corrected density functional theory. The magnitude of corrugation and the shape of the potential energy relief are found to be very sensitive to changes of the shorter wall length at subnanometer scale and atomic structure of the edges if at least one of the walls is chiral. Threshold forces required to start relative motion of the short walls and temperatures at which the transition between

---

[a)] Electronic mail: popov-isan@mail.ru

[b)] Electronic mail: lebedeva@kintechlab.com

[c)] Electronic mail: knizhnik@kintechlab.com

[d)] Electronic mail: lozovik@isan.troitsk.ru

diffusive and free motion of the short walls takes place are estimated. The edges are also shown to provide a considerable contribution to the barrier to relative rotation of commensurate nonchiral walls. For such walls, temperatures of orientational melting, i.e. the crossover from rotational diffusion to free relative rotation, are estimated. The possibility to produce nanotube-based bolt/nut pairs and nanobearings is discussed.

# I. INTRODUCTION

The discovery of relative motion of the walls[1] in multiwalled carbon nanotubes (MWNTs) was immediately followed by the idea that nanotube walls can be used as movable elements[2] of nanoelectromechanical systems (NEMS). In the last decade, a number of NEMS based on the relative motion of nanotube walls have been implemented experimentally. Among these devices, there are nanomotors in which walls of a MWNT play roles of the shaft and bush driven by an electric field[3–5] or a thermal gradient[6] and memory cells operating on relative sliding of the walls along the nanotube axis.[7,8] A wide set of NEMS have been also proposed and studied theoretically, including a gigahertz oscillator,[9,10] a Brownian nanomotor,[11] an accelerometer,[12,13] a nanothermometer,[14,15] an ultrahigh frequency resonator based on the relative vibrations of the nanotube walls,[16] a bolt/nut pair,[6,17–19] a nanoactuator in which a force directed along the nanotube axis leads to rotational motion of the walls[15] and a scanning rotational microscope[20] (see also Ref. 21 for a review). It was also suggested to use the telescopic extension of nanotube walls for thermal nanolithography with an improved spatial resolution.[22]

For some of the applications listed above, such as the nanomotor driven by the electric field, gigahertz oscillator and accelerometer, a free relative motion of nanotube walls is necessary. The operation of some other NEMS that are based on the motion of a short wall relative to a long wall, such as the Brownian nanomotor, nanomotor driven by the thermal gradient, resonator, nanothermometer and bolt/nut pair, is determined by the dependence of interaction energy $U$ of two neighbouring nanotube walls on their relative position. It is convenient to visualize the potential relief of interwall interaction energy $U(x,\varphi)$ as a map plotted on a cylindrical surface, where $x$ is the relative displacement of the walls along the nanotube axis and $\varphi$ is the angle of relative rotation of the walls about this axis. The theoretical analysis[6,17–19,23–34] has shown that there are three types of pairs of adjacent nanotube walls with basically different characteristics of the potential reliefs: (1) commensurate walls at least one of which is chiral, (2) commensurate nonchiral walls (i.e. both armchair or both zigzag walls) and (3) incommensurate walls. For infinite commensurate walls at least one of which is chiral, the potential relief of interwall interaction energy is extremely flat and its corrugations are smaller than the accuracy of calculations.[25,27,28] This is so because incompatibility of helical symmetries of the walls provides that only very high Fourier harmonics of interaction energy $U_a$ between an atom of one of the walls and the second wall contribute to the corrugations of the potential energy relief, whereas the contributions of other harmonics corresponding to different atoms are completely compensated.[24] Contrary to the case of chiral walls, calculations for double-walled nanotubes (DWNTs) with commensurate nonchiral walls performed both in the framework of the density functional theory (DFT)[28–34] and using empirical potentials[13,23–27,34] revealed considerable barriers to relative sliding of the walls along the nanotube axis. However, only two DWNTs with commensurate nonchiral walls having compatible rotational symmetries (5,5)@(10,10)[17,25,27–29,31–33] and (9,0)@(18,0)[25,27,32] were found to have significant barriers to relative rotation of the walls, while the majority of DWNTs with commensurate nonchiral walls have incompatible rotation symmetries of the walls and extremely small barriers to their relative rotation.[27,32] In nanotubes with incommensurate walls, the magnitude of corrugation of the potential energy relief fluctuates near some average value as a function of the overlap length of the walls and, except for these fluctuations, is independent of the system size.[18,19,23]

The results listed above were obtained for infinite systems or without account of the interaction of atoms at the wall edges. However, this interaction can contribute considerably into the corrugations of the potential energy relief. Namely, the extremely flat potential relief for infinite DWNTs with commensurate chiral walls means that in the case of finite DWNTs with no telescopic extension of walls, the potential relief should be determined by the interaction between the lateral surface of the longer wall and atoms at the edges of the shorter wall (since the contributions of these atoms into the total potential relief are not compensated). Analogously, the interaction of atoms at the edges of the shorter wall should also determine the barrier to relative rotation of commensurate nonchiral walls with incompatible rotation symmetries. Furthermore, as the magnitude of corrugation of the potential energy reliefs in DWNTs with incommensurate walls is independent of the system size, it cannot be excluded that in these DWNTs, the interaction of atoms at the edges of the short wall also provides a considerable or even dominant contribution to the total potential energy relief. Thus, the interaction of atoms at wall edges determines the static friction force at the relative motion of chiral commensurate walls and can give a significant contribution to this force for incommensurate walls. Molecular dynamics simulations revealed that the dynamic friction force is also sensitive to the edge structure[35] and the dissipation at the edge atoms can give a dominant contribution to this force.[36] Therefore, account of the interaction of atoms at the wall edges holds the key to the successful development of NEMS based on the relative motion of carbon nanotube walls.

The contribution of edges to the potential relief of interwall interaction energy can be explained in the following way: if the short wall does not consist of an integer number of whole unit cells, contributions of individual atoms of the incomplete unit cell do not compensate each other. The barrier to relative sliding of commensurate (11,2) and (12,12) walls along the axis related to an incomplete unit cell was calculated in paper[23] using the semi-empirical potential. However, the effect of edges on the potential relief of interwall interaction energy is also related to the difference between the interaction of atoms at the shorter wall edges and the lateral surface of the longer wall and the interaction of atoms far from the edges. Due to this difference, contributions of individual atoms to the total potential relief do not compensate each other even for whole unit cells located at wall edges (as opposed to the case of infinite DWNTs with commensurate chiral walls).

A number of experimental studies have been devoted to measurements of physical quantities associated with the relative motion of nanotube walls.[1,6,37,38] The forces needed to pull the core of inner walls out of multi-walled nanotubes were studied.[1,37,38] No overlap-dependent static friction force was revealed.[1,37,38] Instead, considerable irregular but reproducible fluctuations of the pull-out force were observed.[38] Although nanotube walls usually have a helical symmetry, the studies[6] of nanotube-based nanomotors driven by the thermal gradient showed that the relative motion of nanotube walls was helical only rarely. The reasons for the behavior of the pull-out force[38] described above and infrequent occurrence of the helical motion for nanotube-based nanomotors[6] can be understood from the theoretical studies of the effect of wall edges on the potential relief of interwall interaction energy.

In the present paper, we apply the van der Waals corrected density functional theory to calculate the potential reliefs of interwall interaction energy for DWNTs with long inner and short outer walls as well as for DWNTs with both infinite commensurate walls. By comparison of the reliefs for DWNTs with long inner and short outer walls and for DWNTs with both infinite walls in the case when the walls are commensurate and at least one of them is chiral, it is shown that the characteristics of relative motion of such nanotube walls are determined mainly by the contributions of wall edges. The same conclusion is made for DWNTs with incommensurate walls. The dominant contribution of wall edges is also revealed for the barrier to relative rotation of commensurate nonchiral walls. A possible explanation of the experiments[1,6,37,38] mentioned above is provided, dynamical phenomena taking place during the relative motion of the nanotubes walls are considered and the consequences of the results obtained in the present study for design of nanotube-based NEMS are discussed.

The paper is organized in the following way. In Sec. II, the methods used in our calculations are described. In Sec. III, the results on the potential reliefs of interwall interaction energy in DWNTs with different walls are given. The implications of these results for design of nanotube-based NEMS and interpretation of experimental observations on relative sliding of nanotube walls are discussed in Sec. IV.

**II. METHODOLOGY**

The DFT calculations are performed using the VASP code.[39] The Perdew-Wang exchange-correlation functional[40] corrected with the dispersion term[41] is used. The basis set consists of plane waves with the maximum kinetic energy of 500 eV. The interaction of valence electrons with atomic cores is described using Vanderbilt ultrasoft pseudopotentials.[42] A second-order Methfessel–Paxton smearing[43] with a width of 0.1 eV is applied.

The calculations are performed for (4,1)@(12,3), (4,1)@(13,0) and (5,0)@(14,0) nanotubes with finite outer walls and infinite inner walls (see FIG. 1) as well as for (4,1)@(12,3) and (5,0)@(14,0) nanotubes with two infinite walls. Finite outer walls of different length at subnanometer scale and with different atomic structure of the edges are considered (see FIG. 2). The edges of the finite outer walls are terminated with hydrogen atoms. The periodic boundary conditions are applied to model the infinite walls. In calculations for the nanotubes with the (4,1) and (5,0) inner walls and finite outer walls, the sizes of the rectangular model cells along the nanotube axes are 19.52 Å and 17.06 Å, respectively, i.e. the model cells comprise 3 and 4 elementary unit cells of the inner walls. The sizes of the model cells in the directions perpendicular to the nanotube axes equal 17 Å. Integration over the Brillouin zone is performed using the Monkhorst-Pack method[44] with 6 x 1 x 1 and 7 x 1 x 1 k-point grids in the cases of the (4,1) and (5,0) inner walls, respectively. In calculations for the nanotubes with the (4,1) and (5,0) inner walls and infinite (12,3) and (14,0) outer walls, the sizes of the rectangular model cells along the nanotube axes are 6.51 Å and 4.26 Å, respectively, i.e. the model cells comprise one elementary unit cell of the inner walls. For these nanotubes, integration over the Brillouin zone is performed using 18 x 1 x 1 and 30 x 1 x 1 k-point grids, respectively. The structures of the nanotube walls are separately geometrically optimized using the conjugated gradient method until the residual force acting on each atom becomes less than 0.01 eV/Å. Then the walls are shifted along the DWNT axis and rotated relative to each other in order to calculate the potential energy relief. The long-wavelength structural relaxation of commensurate walls of DWNTs due to their interaction has been considered previously within the Frenkel-Kontorova model.[45] Such a relaxation can in principle lead to formation of incommensurability defects corresponding to the relative displacement of the nanotube walls, which can influence the relative motion of the walls. However, this phenomenon can take place for the walls at least hundreds nanometers in length and, thus, is beyond the possibilities of the calculation method used here.

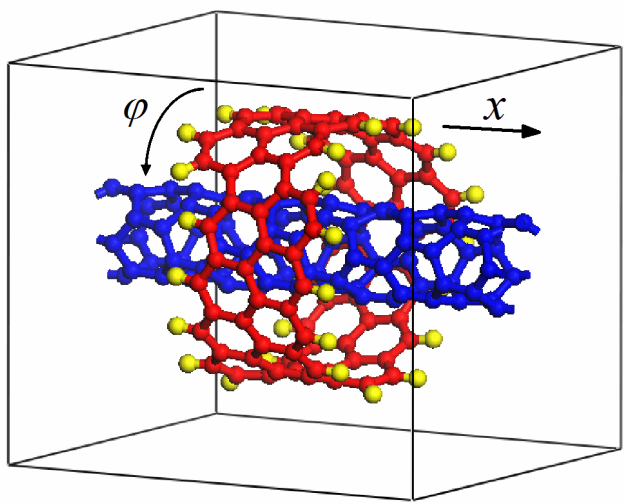

FIG. 1. Model cell with a periodic (4,1) inner wall and a finite (12,3) outer wall terminated with hydrogen atoms. Carbon atoms of the inner and outer walls are coloured in blue and red, respectively. Hydrogen atoms are coloured in yellow.

It should be mentioned that the van der Waals corrected DFT approach has already been used for the analysis of interlayer interaction in bilayer[46–48] and few-layer[49] graphene. The barriers to relative motion of graphene layers calculated in papers[42–44] and paper[48] are in agreement with the value 1.7 ± 0.2 meV/atom estimated[49] on the basis of experimental measurements of shear mode frequencies in few-layer graphene and graphite[50] and of static friction between the nanometer-sized graphene flake and graphite surface using the friction force microscope[51] within 20% and 70%, respectively.

As for the empirical Lennard-Jones potential (which was widely applied for studies of the relative motion of nanotubes walls,[9–11,13,17–19,22,26,27] the barrier to relative motion of graphene layers calculated using this potential is underestimated by more than an order of magnitude. Moreover, only *ab initio* methods are able to describe the difference in the interaction between edge atoms of graphene-like layers and atoms far from the edges, whereas empirical potentials do not distinguish such a difference. However, as the interaction with the edge atoms can considerably contribute to the total potential relief of interwall interaction energy, the adequate account of this difference is essential for the problem under consideration. Thus, we believe that the van der Waals corrected DFT approach used in the present paper is appropriate for qualitative studies of the influence of edges on the relative motion of walls in carbon nanotubes.

## III.   RESULTS

Let us first consider potential reliefs of interaction energy between commensurate chiral walls of the (4,1)@(12,3) DWNT. In the case when both the walls are infinite, the potential relief of interlayer interaction energy is found to be extremely smooth, with the magnitude of corrugation below 0.1 meV per unit cell. Thus, we confirm the results of previous DFT calculations without the van der Waals correction[28] and calculations using empirical potentials.[23,25,27] The result obtained means that an addition of an integer number of whole unit cells to the short wall can lead to changes of the potential relief only within 0.1 meV per unit cell (the whole unit cells of the short outer walls considered are indicated in FIG. 2).

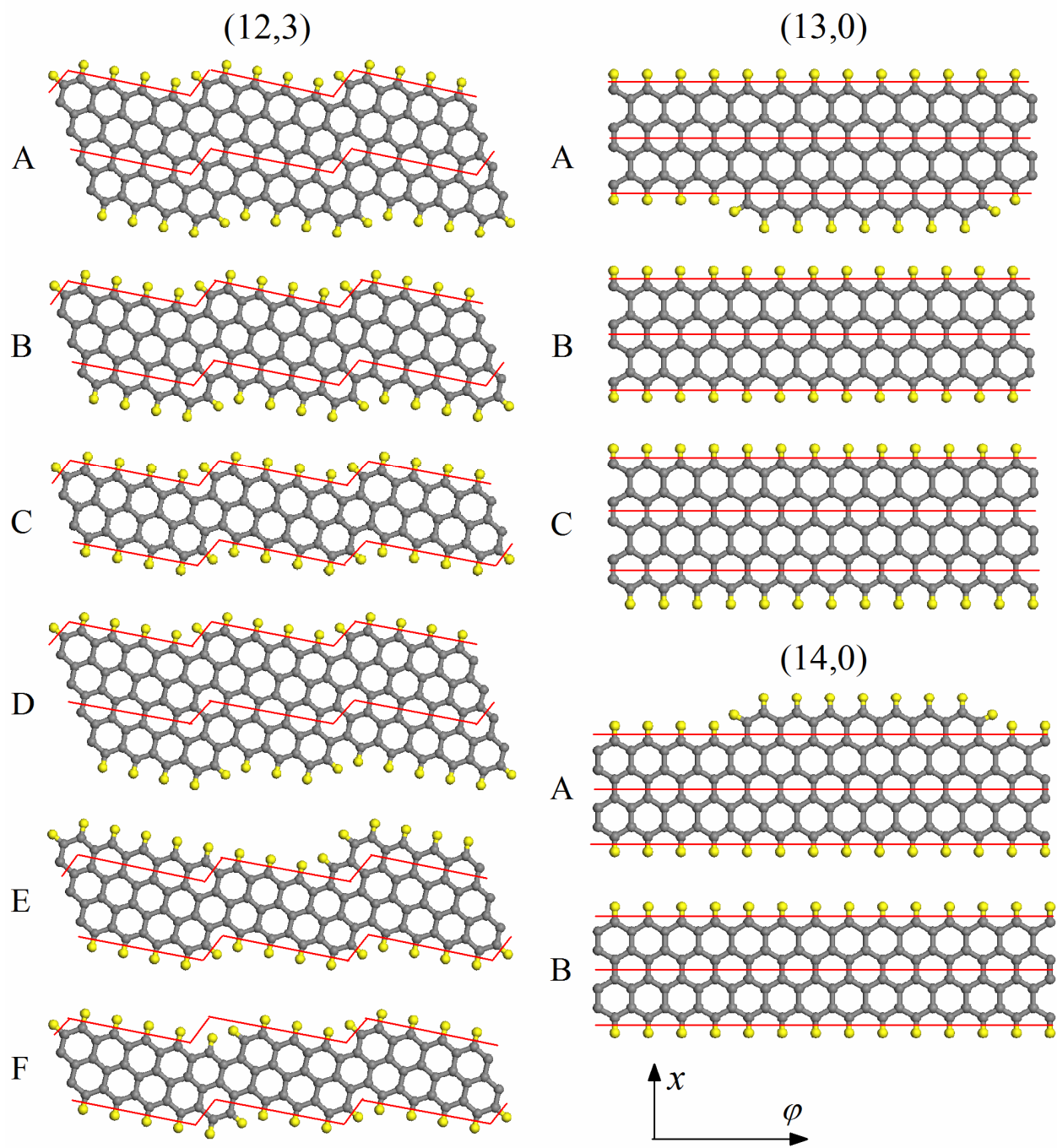

FIG. 2. Considered outer walls (12,3), (13,0) and (14,0) unfolded onto a plane. Carbon and hydrogen atoms are coloured in gray and yellow, respectively. Unit cells of the outer walls are indicated by red solid lines.

For the (4,1)@(12,3) DWNTs with the finite outer walls containing *incomplete unit cells* and having different length and edge structure, the magnitude of corrugation of the potential energy relief is found to range from 3 to 55 meV (see TABLE I). Comparing these values with the upper bound for the infinite outer wall, it can be estimated that the edges provide a dominant contribution to the potential energy relief for outer wall lengths up to at least 20 – 300 nm.

The basically different shapes of the potential reliefs are revealed for the (4,1)@(12,3) DWNTs with the finite outer walls that have and do not have the rotational symmetry $C_3$. In the cases when the finite outer wall has a rotational symmetry, the potential relief of interwall interaction energy has a prominent thread-like shape with a negligible barrier to relative motion of the walls along the thread and a high barrier to relative motion of the walls across the thread (see FIG. 3a-c). On the graphene surface formed by unfolding the inner wall, the angle of the thread line with respect to the wall circumference equals $90º - \alpha$, where $\alpha = 11º$ is the chiral angle of the (4,1) inner wall. This means that the thread line is parallel to the armchair direction. It should be noted that thread-like potential energy reliefs are also typical for DWNTs with commensurate chiral walls both of which are infinite and one of which contains defects at identical positions for a number of nanotube unit cells.[27,52,53] However, in those systems, the thread line is parallel to the zigzag direction on the graphene surface formed by unfolding the perfect inner or outer wall. The height of the barrier to relative motion of the short outer wall across the thread is found to depend strongly on the length of the shorter wall part corresponding to the incomplete unit cell. Namely, a removal or an addition of just one row of carbon atoms at the edge of the (12,3) wall (see structures (12,3)A – C in FIG. 2) can result in the change of the barrier by an order of magnitude (see FIG. 3a–c).

TABLE I. Calculated magnitudes $E_{max}$ of corrugation of the potential reliefs of interwall interaction energy in different DWNTs, barriers $E_1$ and $E_2$ to relative motion of nanotube walls along and across the thread lines, threshold forces $F_1$ and $F_2$ required to start the relative motion of the short outer walls from the global energy minima along and across the thread lines. The numbers $N_C$ and $N_H$ of carbon and hydrogen atoms in the outer walls (per unit cell in the cases of the infinite outer walls) and rotational symmetries of the outer walls are indicated. The structures of the outer walls are shown in FIG. 2.

| Outer walls | $N_C$ | $N_H$ | Rotational symmetry | $E_{max}$ (meV) | $E_1$ (meV) | $E_2$ (meV) | $F_1$ (pN) | $F_2$ (pN) |
|---|---|---|---|---|---|---|---|---|
| infinite inner (4,1) wall and different outer walls ||||||||||
| infinite (12,3) | 84 | - | $C_3$ | <0.1 | | | | |
| (12,3)A | 144 | 30 | $C_3$ | 21.4 ± 0.8 | <0.8 | 21.4 ± 0.8 | <1.2 | 33.1 ± 1.2 |
| (12,3)B | 114 | 30 | $C_3$ | 2.8 ± 0.8 | <0.8 | 2.8 ± 0.8 | <1.2 | 4.2 ± 1.2 |
| (12,3)C | 84 | 30 | $C_3$ | 54.6 ± 1.0 | < 1.0 | 54.6 ± 1.0 | <1.4 | 84.4 ± 1.5 |
| (12,3)D | 142 | 30 | $C_1$ | 34.6 ± 0.8 | 16.0 ± 0.8 | 24.1 ± 0.8 | 19 ± 4, 17 ± 4 | 32 ± 8, 30 ± 3 |
| (12,3)E | 106 | 30 | $C_1$ | 16.8 ± 1.6 | 7.3 ± 1.6 | 8.5 ± 1.6 | 10 ± 3, 7.5 ± 1.6 | 12 ± 2, 9.6 ± 0.5 |
| (12,3)F | 84 | 30 | $C_1$ | 48.1 ± 0.4 | 4.8 ± 0.5 | 20.3 ± 0.6 | 10 ± 4 | 26 ± 3 |
| (13,0)A | 119 | 27 | $C_1$ | 49.1 ± 0.2 | 15.4 ± 0.4, 19.5 ± 0.5 | 28.9 ± 0.8, 33.1 ± 0.6 | 27 ± 2, 34 ± 5 | 37 ± 3, 47 ± 6 |
| (13,0)B | 104 | 26 | $C_{13}$ | <1.3 | | | | |
| (13,0)C | 130 | 26 | $C_{13}$ | <5.5 | | | | |
| infinite (5,0) wall and different outer walls ||||||||||
| infinite (14,0) | 56 | - | $C_{14}$ | 43.97 ± 0.08 | <0.08 | 43.97 ± 0.08 | <0.07 | 103.9 ± 0.2 |
| (14,0)A | 127 | 29 | $C_1$ | 104.4 ± 1.7 | 12.6 ± 1.7 | 93.3 ± 1.7 | 11 ± 2 | 223 ± 2 |
| (14,0)B | 112 | 28 | $C_{14}$ | 82.1 ± 0.7 | <0.23 | 82 ± 0.7 | <0.20 | 194 ± 3 |

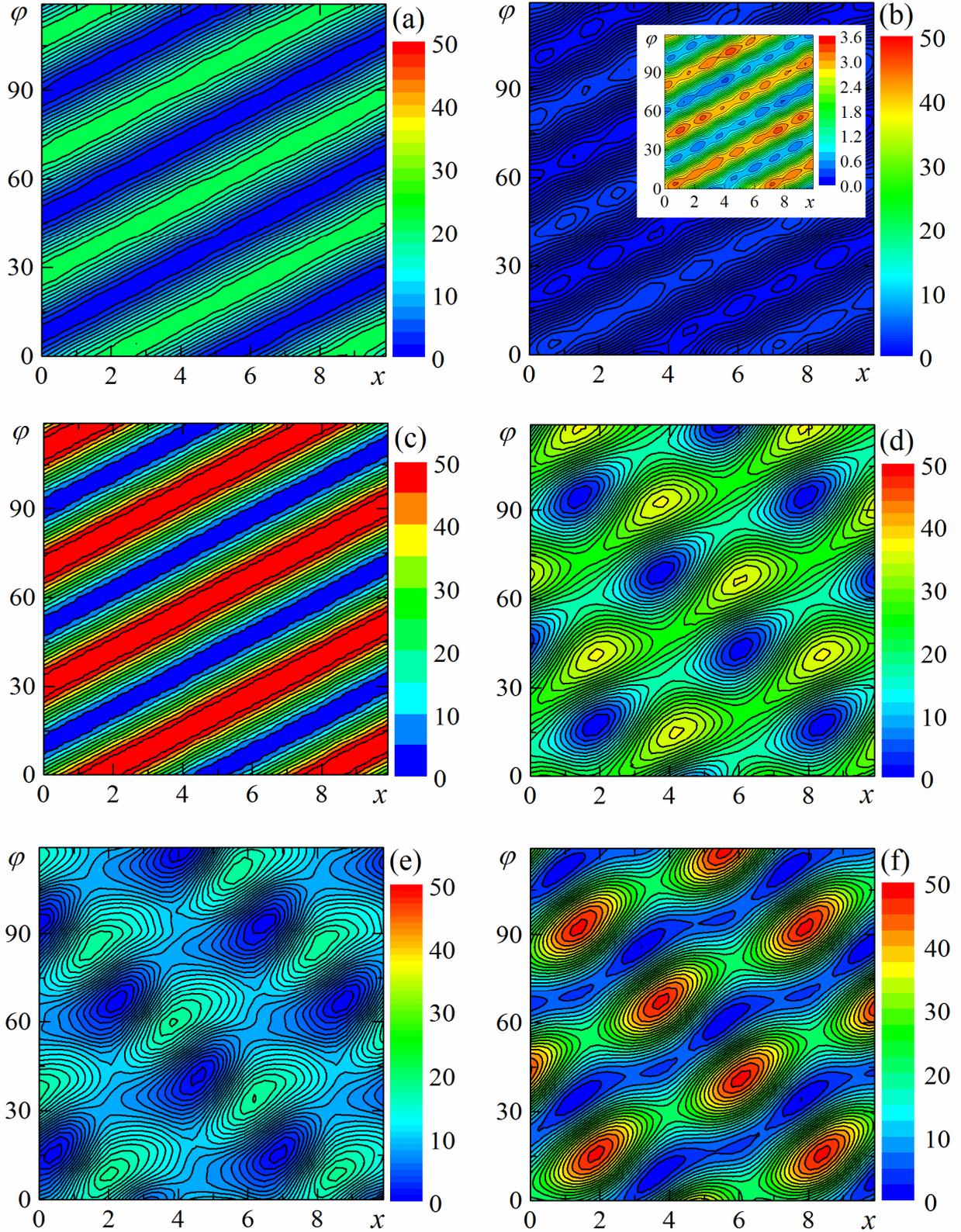

FIG. 3. Calculated interwall interaction energies (in meV) of the (4,1)@(12,3) DWNTs with the infinite inner wall and finite outer walls as functions of the relative displacement $x$ (in Å) of the walls along the nanotube axis and the angle $\varphi$ (in degrees) of relative rotation of the walls about this axis. The order of the potential reliefs corresponds to the order of structures of the (12,3) A – F outer walls in FIG. 2. The equipotential lines are drawn with steps (a, d) 2.0 meV, (b) 0.2 meV, (c) 5.0 meV, (e) 1.0 meV, and (f) 2.5 meV.

In the cases when the outer wall does not have a rotational symmetry, there is not only the barrier to relative motion of the walls across the thread line but also a considerable barrier to relative motion of the walls along the thread line (see FIG. 3d-f). Thus, breaking the rotational symmetry by introduction of just an additional couple of atoms leads to the drastic change of the shape of the potential energy relief (see FIG. 3d). Two types of minima, a deep one and a small one, can be seen on some potential energy reliefs. For example, two types of energy minima with the energy difference of $4.0 \pm 0.4$ meV are found for the (12,3)F wall (see FIG. 3f). However, the small minima become completely unstable for other structures of the outer wall edges (see FIG. 3d and e).

The calculations for the incommensurate (4,1)@(13,0) DWNTs with infinite inner and finite outer walls also reveal a relation between the rotational symmetry of the short outer wall and characteristics of the potential relief of interwall interaction energy. Namely, the smooth potential reliefs are found for the short walls of high rotational symmetry. The magnitude of corrugation of the potential relief in this case (below 6 meV) is the smallest among all the considered DWNTs with finite outer walls (see TABLE I). However, when the rotational symmetry of the outer wall is broken by roughness of the edges, the magnitude of corrugation of the potential energy relief is observed to increase almost to 50 meV (see FIG. 4). The barriers to relative motion of the walls along and across the thread line in this case are found to be comparable. Moreover, two types of prominent minima and maxima with the energy difference $5.1 \pm 0.2$ meV are observed on the potential relief. We believe that a high rotational symmetry leads to the compensation of contributions of individual atoms into the total potential relief. A sufficiently high rotational symmetry of the finite wall (e.g., of 13-th order in the considered case) is possible only for nonchiral walls with smooth edges. Thus, we propose that pairs of long chiral and short nonchiral walls with smooth edges can be used in NEMS (for example, nanomotors) in which the small static friction force for relative motion of the walls is necessary.

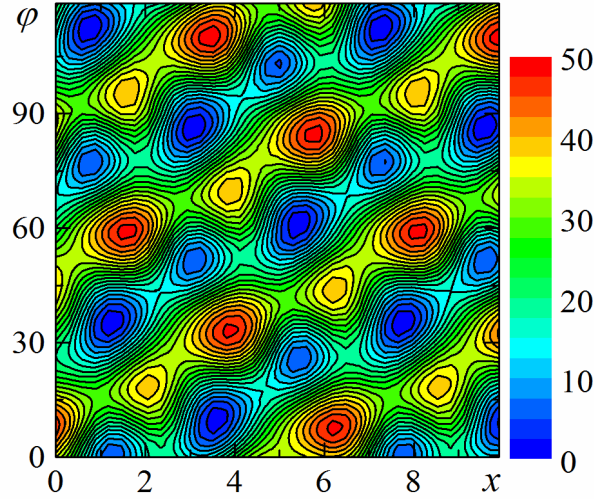

FIG. 4. Calculated interwall interaction energy (in meV) of the (4,1)@(13,0) DWNT with the infinite inner wall and finite (13,0)A outer wall as a function of the relative displacement $x$ (in Å) of the walls along the nanotube axis and the angle $\varphi$ (in degrees) of relative rotation of the walls about this axis. The structure of the finite outer wall is shown in FIG. 2. The equipotential lines are drawn with a step 2.5 meV.

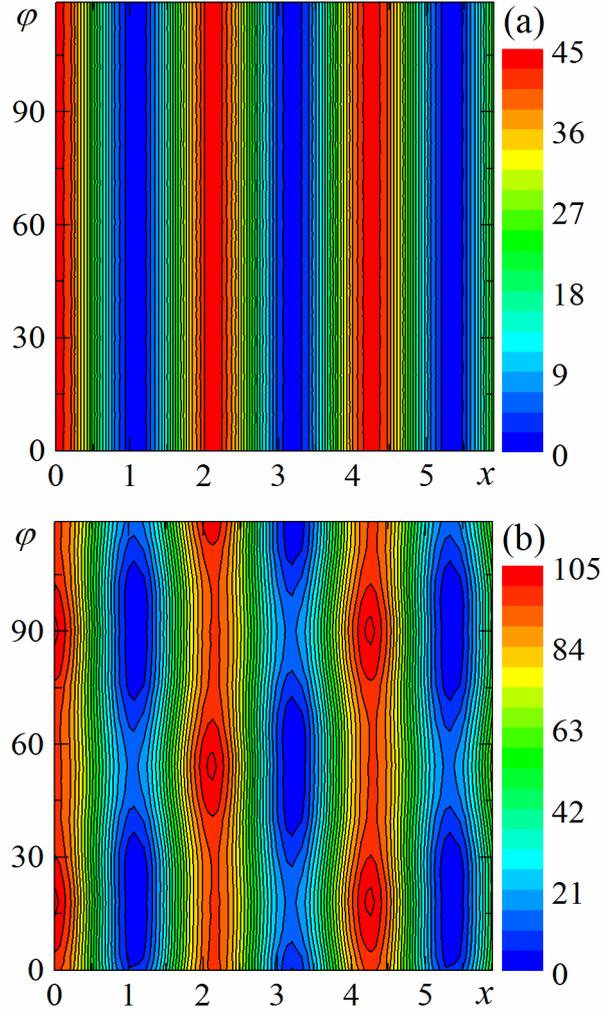

FIG. 5. Calculated interwall interaction energies (in meV) of the (5,0)@(14,0) DWNTs as functions of the relative displacement $x$ (in Å) of the walls along the nanotube axis and the angle $\varphi$ (in degrees) of relative rotation of the walls about this axis: (a) for one unit cell of the DWNT with both infinite walls, (b) for the DWNT with the infinite inner wall and finite outer (14,0)A wall. The structure of the finite outer wall is shown in FIG. 2. The equipotential lines are drawn with steps (a) 2.22 meV and (b) 5.25 meV.

The results described above for the DWNTs with commensurate walls at least one of which is chiral and with incommensurate walls show that for both these types of DWNTs, the contribution of the wall edges to the potential energy relief can be dominant. As opposed to these DWNTs, the barrier to relative motion of commensurate nonchiral walls along the DWNT axis scales linearly with the overlap length of the walls[13,23–34] and, therefore, the contribution of the wall edges to this barrier should generally be small. For example, for the (5,0)@(14,0) DWNT, the barrier to relative motion of the nanotube walls is calculated to be 44 meV per unit cell (see FIG. 5a) and, therefore, the total barrier considerably exceeds the maximal observed corrugation of the potential energy relief induced by the wall edges ~50 meV already for several unit cells. However, due to incompatibility of rotational symmetries of walls in most DWNTs with commensurate nonchiral walls, the barriers to relative rotation of the infinite walls about their common axis are negligibly small. For the (5,0)@(14,0) DWNT with both infinite walls we estimate this barrier to be below 0.08 meV per unit cell (see FIG. 5a). The barrier to relative rotation of the walls is still found to be small (<0.23 meV for the wall consisting of 2 whole unit cells) if the outer wall has the edges preserving the rotational symmetry. However, once the rotational symmetry is broken by roughness of the walls edges, a barrier of about 12.6 meV appears (see FIG. 5b). Thus, at least for relatively short outer walls of the (5,0)@(14,0) DWNT with the length below 70 nm, the contribution of the rough edges of the short wall to the barrier to relative rotation of the walls is dominant.

Since for infinite DWNTs with commensurate nonchiral walls having incompatible rotational symmetries, the barrier to relative rotation of the walls is several orders of magnitude smaller than the barrier to relative sliding of the walls along the nanotube axis, such nanotubes were proposed to be used as easy-to-operate nanobearings with a fixed position of the bush based on the outer wall along the axis.[27] The calculations performed here show that the high ratio of these barriers can be achieved only in the case of smooth edges, when the short wall has a high rotational symmetry.

Numerous first-principles and empirical calculations have demonstrated recently that potential reliefs of interaction energy of infinite graphene layers[46–49,51] and infinite commensurate nanotube walls in the cases when both of the walls are nonchiral[16,20,27,28,33] or both of the walls are chiral and one of them contains defects at identical positions for a number of nanotube unit cells[20,53] can be approximated with high accuracy by expressions containing only the first Fourier harmonics. However, our calculations for the DWNTs with the finite outer walls show that in such systems, a variety of complex potential reliefs of interwall interaction energy is observed depending on the edge structure. In general, these potential energy reliefs cannot be fitted with a simple approximation containing only the first Fourier harmonics (see FIG. 3 – FIG. 5). An exception is the case when the outer wall has a rotational symmetry. In such systems, the potential energy reliefs can be fitted closely with just one harmonic term. For example, the potential reliefs of interaction energy between the infinite (4,1) inner wall and finite (12,3)A and (12,3)C outer walls (see FIG. 2) can be approximated by the expression

$$U(x,\varphi) = U_0 + \frac{E_2}{2}\left[1 - \cos\left(kx\cos\left(\frac{\pi}{2} - \alpha\right) - kr\varphi\sin\left(\frac{\pi}{2} - \alpha\right)\right)\right], \qquad (1)$$

where $k = 4\pi/(\sqrt{3}l)$, $l = 1.42$ Å is the nanotube bond length and $r = 1.8$ Å is the radius of the perfect (4,1) inner wall. It should be noted that this expression is not exactly similar to the ones used in papers[20,53] for DWNTs with commensurate chiral walls both of which are infinite and one of which contains defects at identical positions for a number of nanotube unit cells. First, this expression contains only one harmonic term corresponding to variation of the interwall interaction energy with the displacement across the thread line, while in papers,[20,53] both the variations along and across the thread line were included. Second, the thread angle and the distance between the thread lines in expression (1) are different from those in papers.[20,53] As mentioned above, for the (4,1)@(12,3) DWNTs with finite outer walls, the thread line is parallel to the armchair direction on the graphene surface formed by unfolding the inner wall, while for DWNTs with commensurate chiral walls both of which are infinite and one of which contains defects at identical positions for a number of nanotube unit cells, the thread line is parallel to the zigzag direction. The root-mean-square deviations of approximations (1) with $E_2 = 21.4$ meV and $E_2 = 54.6$ meV from the potential energy reliefs calculated for (4,1)@(12,3) DWNTs with the (12,3)A and (12,3)C outer walls (see FIG. 2) are found to be only 0.42 meV and 0.37 meV, respectively.

The calculated potential reliefs of interwall interaction energy in the DWNTs can be used to estimate threshold forces required to start motion of the short outer wall relative the fixed inner wall (see TABLE I). The forces acting on each atom of the outer wall are assumed to have the form $\vec{F} = \vec{F}_x + \vec{F}_R$, where components $\vec{F}_x$ and $\vec{F}_R$ are directed along the nanotube axis and along the tangent to the outer wall circumference, respectively, and are equal in magnitude for all the atoms, so that no deformations of the walls are induced. Thus, the threshold forces are determined as the maximal gradients of interwall interaction energy along the minimum energy paths between adjacent energy minima on the potential energy relief $U(x, R\varphi)$, where $R$ is the outer wall radius. The estimated threshold forces $F_1$ and $F_2$ required to start the relative motion of the outer walls along and across the thread lines are listed in TABLE I. For all the considered DWNTs, contributions of the outer wall edges to the threshold forces for relative motion of the walls are found to be within 100 pN. These forces are comparable to the fluctuations of interwall interaction force observed in the experiment based on the telescopic extension of walls of greater radii and undetermined chirality indices[38] (~100 pN).

The calculated potential reliefs also allow evaluating the dynamics of relative motion of the nanotube walls. While in DWNTs with both infinite walls at least one of which is chiral, the barriers to relative motion of the walls along the axis are negligible and it can be expected that such a motion is free down to very low temperatures, the presence of edges without a rotational symmetry provides a corrugated potential energy relief and, therefore, impeded relative motion of the walls at low temperatures. The diffusion coefficients of the short outer walls along the thread line at temperatures $k_B T \ll E_1$ can be estimated as[18,19]

$$D = \frac{\Omega \delta^2}{2} \exp\left(-\frac{E_1}{k_B T}\right), \qquad (2)$$

where $k_B$ is the Boltzmann constant, $\delta = 3l \cos\alpha$ is the distance between adjacent energy minima along the nanotube axis and $\Omega$ is the frequency of relative vibrations of the walls about the energy minima. For the considered DWNTs with the inner (4,1) wall, we estimate the temperature $T \sim E_1 / k_B$ below which the relative motion of the walls is diffusive to be 190 K, 180 K, 85 K and 56 K for the (12,3)D, (13,0)A, (12,3)E and (12,3)F outer walls (see FIG. 2), respectively. Nevertheless, even with account of the effect of wall edges, the relative motion of the outer walls in the considered DWNTs along the thread lines is found to be free at room temperature. This conclusion is in qualitative agreement with the fact that the free motion along the DWNT axis was observed at room temperature for the short inner wall of radius 4.7 Å.[54] Taking $\delta = 2\pi / n$ for DWNTs with commensurate nonchiral walls, the same expression (2) can be used to estimate the rotational diffusion coefficient of the short wall about the ($n$, 0) or ($n$, $n$) long wall.

For infinite DWNTs with commensurate nonchiral walls, the phenomenon of orientational melting (i.e. the transition between the free motion and diffusion regimes) was proposed.[27,32] As the interaction between commensurate nonchiral walls having incompatible rotational symmetries does not contribute to the barrier $E_1$ to their relative rotation if these walls are infinite, perfect and rigid,[27,32] such a barrier can arise due to the following reasons: (1) edge effect related to the interaction between the lateral surface of one of walls and the edge of the other wall; (2) deformations of the walls due to thermodynamic fluctuations and interwall interaction; (3) distortions of the wall structure related to spontaneous symmetry breaking. Here we consider only the first of the reasons listed above and calculate the barrier $E_1$ which originates from the edge effect. If the calculated edge-related barrier $E_1$ for the (5,0)@(14,0) DWNT gives the main contribution into the real barrier to relative rotation of the walls, the temperature of orientational melting can be estimated as $T \sim E_1/k_B$ and is found to be from below 3 K for the (14,0)B outer wall with the smooth edges of high rotational symmetry to 150 K for the (14,0)A outer wall with the rough edges without rotational symmetry. The molecular dynamics simulations of orientational melting of a double-shell carbon nanoparticle with the temperature of orientational melting about 100 K showed that thermally activated deformations of the shells do not contribute to the average barrier to relative rotation of the shells at this temperature.[55,56] As for spontaneous symmetry breaking, related distortions of the nanotube structure can arise at temperatures below the temperature of Peierls transition, which for armchair nanotubes with the diameter greater than 1 nm, was estimated to be less than 1 K,[57] about 10 K,[58,59] 15 K,[60] and 40 K.[61] These temperatures are considerably lower than the estimated temperature of orientational melting in the case of rough edges. Thus, we assume that in the case of sufficiently short walls with rough edges without rotational symmetry, the temperature of orientational melting is determined by the edge effect. However, for long walls, the influence of long-wavelength bending and torsional modes on the barrier to relative rotation of the walls may also become important. It should be mentioned that a clear anomaly in the temperature dependence of specific heat of multi-walled nanotubes at 60 K was attributed to melting of orientational dislocations.[62] In the case of walls with rotational symmetry, the contribution of edges into the barrier to relative rotation of the walls is negligible and, therefore, the temperature of orientational melting should be determined by other reasons such as spontaneous symmetry breaking and deformations of the walls due to thermodynamic fluctuations including excitation of long-wavelength bending and torsional modes.

## IV. DISCUSSION AND CONCLUSIONS

The van der Waals corrected density functional theory has been applied to study the interwall interaction energy in DWNTs with infinite inner walls and short outer walls as a function of the angle of relative rotation of walls about the DWNT axis and their relative displacement along the axis. The calculations performed show that for DWNTs with incommensurate walls or with commensurate walls at least one of which is chiral not only the magnitude of corrugation of the potential relief but also the qualitative shape of the relief are very sensitive both to the structure of wall edges and to changes of the shorter wall length at subnanometer scale. Namely, we have found that in DWNTs with the same chiral indices but with the different structure of wall edges (for the same length of the short wall) or slightly different length of the short wall, the potential relief of interwall interaction energy can have two qualitatively different shapes: 1) the thread-like shape with the barrier to motion of the short wall along the helical thread line that is an order of magnitude smaller than the barrier to motion across the thread line and 2) the relief with comparable values of the barriers between all adjacent minima in different directions. It should be noted that this conclusion is valid for any length of the short wall in the case of incommensurate walls and up to the short wall length of at least 20 nm in the case of commensurate walls at least one of which is chiral.

The significant contribution of the wall edges to the potential relief of interwall interaction energy is revealed not only for DWNTs with commensurate chiral walls and incommensurate walls but also for DWNTs with commensurate nonchiral walls. For the (5,0)@(14,0) DWNT, the edges of the short wall with no rotational symmetry are expected to provide the dominant contribution to the barrier to relative rotation for the short wall lengths up to at least 70 nm. The temperature of orientational melting is estimated for this DWNT to be from below 3 K for the outer wall with the smooth edges of high rotational symmetry to 150 K for the outer wall with the rough edges without rotational symmetry.

Let us discuss the conclusions that can be made on the basis of the calculated potential reliefs for the experimental studies of relative motion of nanotubes walls.[1,6,38] The contributions of edges to threshold forces for relative motion of the nanotube walls are found to be within 100 pN for the considered DWNTs of diameter about 1 nm. This is on the order of static friction forces measured in the high-resolution experiment[38] for nanotubes of diameter about 5 nm (~100 pN). As a random pair of neighbour walls should be incommensurate in the majority of cases, such a type of interacting walls is usually considered experimentally. In the experiment,[38] it was revealed that at telescopic extension of the core of inner walls, 1) the overlap-dependent contribution to the static friction force (equal to the force that is necessary for relative motion of walls) was negligibly small, and 2) this force exhibited irregular but reproducible fluctuations. In the case of telescopic extension, the edges of both neighbor walls contribute to the total potential relief and the related dependence of the static friction force on the wall overlap. The translational period of the contributions corresponding to each edge is determined by the translational period of the neighbor wall. If the walls are incommensurate, the superposition of the contributions from the inner and outer wall edges leads to the irregular total dependence of the static friction force on the wall overlap. Thus, the potential reliefs calculated with account of the effect of wall edges allow to explain not only experimentally measured magnitudes of the static friction force but also the behavior of this force with changing the overlap length of the walls.

While thermally activated rotation of short outer walls was observed recently for a set of nanotubes, the helical motion of the wall was revealed only in seldom cases.[6] We have found that the prominent thread-like potential relief and, therefore, the helical motion of the shorter outer wall are possible only for the certain lengths and edge structures of the wall. This can be the reason why the helical motion was rarely observed in the experiment.[6]

The strong influence of edge structure on the potential relief of interwall interaction energy is of great importance for the possibility to produce nanotube-based nanobearings and bolt/nut pairs. The DWNTs with commensurate nonchiral walls having incompatible rotational symmetries were proposed to be used as easy-to-operate nanobearings with a fixed position of the outer wall-based bush along the axis. The calculations performed show that this is possible only in the case of the short wall with smooth edges. It was shown recently that reconstruction of graphene edges under electron irradiation in the transmission electron microscope leads to the formation of smooth zigzag edges.[63] Since hydrogen is easily removed under the conditions when the irradiation-induced transformations of graphene-like carbon network take place, the formation of smooth edges can be related to the decrease in the number of dangling bonds. For nonchiral walls, smooth edges have the minimal number of dangling bonds. Thus, we propose that nonchiral walls with smooth edges can be obtained by exposure of the walls to electron irradiation or thermal annealing.

A high ratio of the barriers to motion of the movable wall along and across the helical thread line (a so-called relative thread depth[18,19]) is essential for the use of a DWNT as an easy-to-operate bolt/nut pair. The present study shows that to achieve such values of the relative thread depth it is not sufficient just to choose the walls with certain chiral indices. The control of the wall length and edge structure *with atomic precision* is also necessary. Unfortunately, the production of DWNTs with pre-determined chiral indices of both walls and controlled lengths and edge structure of shorter walls is beyond the possibilities of modern nanotechnology. For chiral walls, edges with rotational symmetry do not have the minimal number of dangling bonds. Thus, such edges cannot be produced by electron irradiation as proposed above for nonchiral walls. So in order to obtain nanotube-based bolt/nut pairs at the moment it is necessary to sort out tens of arbitrary nanotubes and to test the possibility of helical motion of short outer walls (for example, by studies of their diffusion). However, in this case, the characteristics of the thread may not reach the desired values. The only possible way to get nanotube-based bolt/nut pairs with given characteristics of the thread that has been proposed so far is the creation of identical atomic scale defects (for example, vacancies) at identical positions in many unit cells of short walls.[27,52,53]

ACKNOWLEDGEMENTS

This work was supported by the RFBR grants 11-02-00604 and 12-02-90041-Bel. The calculations were performed on the SKIF MSU Chebyshev supercomputer, the MVS-100K supercomputer at the Joint Supercomputer Center of the Russian Academy of Sciences and the Multipurpose Computing Complex NRC "Kurchatov Institute".